\documentclass[aip, amsmath, amssymb, reprint, floatfix]{revtex4-2}

\usepackage{graphicx}
\usepackage{dcolumn}
\usepackage{bm}

\usepackage[utf8]{inputenc}
\usepackage[T1]{fontenc}
\usepackage{mathptmx}
\usepackage{siunitx}
\usepackage{xcolor}
\colorlet{MyChanges}{cyan!70!blue!80!}

\begin{document}

\title{Fabrication of Sawfish photonic crystal cavities in bulk diamond}

\author{Tommaso Pregnolato}
\affiliation{Ferdinand-Braun-Institut gGmbH, Leibniz-Institut f\"{u}r H\"{o}chstfrequenztechnik, Gustav-Kirchhoff-Str. 4, 12489 Berlin, Germany}
\affiliation{Department of Physics, Humboldt-Universit\"{a}t zu Berlin, Newtonstr. 15, 12489 Berlin, Germany}

\author{Marco E. Stucki}%
\affiliation{Ferdinand-Braun-Institut gGmbH, Leibniz-Institut f\"{u}r H\"{o}chstfrequenztechnik, Gustav-Kirchhoff-Str. 4, 12489 Berlin, Germany}
\affiliation{Department of Physics, Humboldt-Universit\"{a}t zu Berlin, Newtonstr. 15, 12489 Berlin, Germany}

\author{Julian M. Bopp}%
\affiliation{Ferdinand-Braun-Institut gGmbH, Leibniz-Institut f\"{u}r H\"{o}chstfrequenztechnik, Gustav-Kirchhoff-Str. 4, 12489 Berlin, Germany}
\affiliation{Department of Physics, Humboldt-Universit\"{a}t zu Berlin, Newtonstr. 15, 12489 Berlin, Germany}

\author{Maarten H. v. d. Hoeven}%
\affiliation{Department of Physics, Humboldt-Universit\"{a}t zu Berlin, Newtonstr. 15, 12489 Berlin, Germany}

\author{Alok Gokhale}%
\affiliation{Department of Physics, Humboldt-Universit\"{a}t zu Berlin, Newtonstr. 15, 12489 Berlin, Germany}

\author{Olaf Kr\"{u}ger}%
\affiliation{Ferdinand-Braun-Institut gGmbH, Leibniz-Institut f\"{u}r H\"{o}chstfrequenztechnik, Gustav-Kirchhoff-Str. 4, 12489 Berlin, Germany}

\author{Tim Schr\"{o}der}%
\email{tim.schroeder@physik.hu-berlin.de}
\affiliation{Ferdinand-Braun-Institut gGmbH, Leibniz-Institut f\"{u}r H\"{o}chstfrequenztechnik, Gustav-Kirchhoff-Str. 4, 12489 Berlin, Germany}
\affiliation{Department of Physics, Humboldt-Universit\"{a}t zu Berlin, Newtonstr. 15, 12489 Berlin, Germany}

\date{\today}

\begin{abstract} 
Color centers in diamond are quantum systems with optically active spin-states that show long coherence times and are therefore a promising candidate for the development of efficient spin-photon interfaces. However, only a small portion of the emitted photons is generated by the coherent optical transition of the zero-phonon line (ZPL), which limits the overall performance of the system. Embedding these emitters in photonic crystal cavities improves the coupling to the ZPL photons and increases their emission rate. Here, we demonstrate the fabrication process of "Sawfish" cavities, a design recently proposed that has the experimentally-realistic potential to simultaneously enhance the emission rate by a factor of 46 and couple photons into a single-mode fiber with an efficiency of 88\%. The presented process allows for the fabrication of fully suspended devices with a total length of \SI{20.5}{\micro\meter} and  features size as small as \SI{40}{\nano\meter}. The optical characterization shows fundamental mode resonances that follow the behavior expected from the corresponding design parameters and quality (Q) factors as high as 3825. Finally, we investigate the effects of nanofabrication on the devices and show that, despite a noticeable erosion of the fine features, the measured cavity resonances deviate by only 0.9 (1.2)\% from the corresponding simulated values. This proves that the Sawfish design is robust against fabrication imperfections, which makes it an attractive choice for the development of quantum photonic networks.
\end{abstract}

\maketitle

\section{Introduction}
Quantum-photonic networks are envisioned as formed by distributed nodes that host optically active quantum systems with long coherence times and that are interconnected via photonic channels. Such systems require an efficient light-matter interface, which enables the coupling between the nodes and the propagating photons.~\cite{Atature2018} Color centers in diamond are promising candidates for building such a platform, because these solid-state quantum systems have optically-addressable and coherent spin states.~\cite{Atature2018, Bradac2019, Schroder2016} The negatively charged nitrogen vacancy (NV$^{-}$) center, for example, has already been employed for the generation of entanglement between distant quantum systems.~\cite{Hensen2015, Humphreys2018, Pompili2021} The performance of this system, however, relies on how many of the emitted photons are generated by the coherent ZPL transition. This ratio is quantified by the Debye-Waller factor and for NV$^{-}$ it amounts to only 3$\%$ of the total emission.~\cite{Atature2018} The emerging group-IV defects have shown larger coherent emission of up to 80\%,~\cite{Neu2011} but they still present Debye-Waller factors that are far from unity.~\cite{Bradac2019, Gorlitz2020}

In order to further improve the emission into the ZPL, many works have focused on the development and fabrication of photonic crystal cavities (PhCCs) in diamond.~\cite{Li2015, Schroder2016, Latawiec2016a, Mouradian2017, Kuruma2021, Rugar2021, Bhaskar2020} Characterized by structures that periodically modulate the refractive index of the substrate, PhCCs modify the local density of optical states surrounding the emitter and thus can enhance the coupling between the spin levels of the color centers and the interacting photons at the position of the increased density of states. Such an effect can be measured as a spectrally resolved increased emission rate of the ZPL photons and it is quantified by a Purcell factor $F_{\text{p}}>1$.~\cite{Purcell1946}

Recently, we proposed a new design for a 1-dimensional PhCC, named "Sawfish" cavity,~\cite{Bopp2022} which has the potential to simultaneously achieve $F_{\text{p}}=46$ and 88\% coupling efficiency into a single-mode fiber for the ZPL photons emitted by an embedded negatively-charged tin vacancy center (SnV$^-$). Our simulations show that such performance can be already attained for a device designed with realistic parameters and presenting a Q of 17000.~\cite{Bopp2022} However, the Sawfish design requires a fully suspended nanostructure in order to efficiently confine the light. The fabrication of fully free-stranding nanostructures is routinely performed in many material platforms, e.g. III-V-semiconductor heterostructures, where wet etching steps enable the selective removal of specific layers without affecting others.~\cite{Midolo2015} Conversely, the high chemical stability of diamond hinders an approach based on wet chemistry and requires different methods for the production of suspended nanodevices. 

\begin{figure}[b!]
\begin{center}
\includegraphics[width=8.5cm]{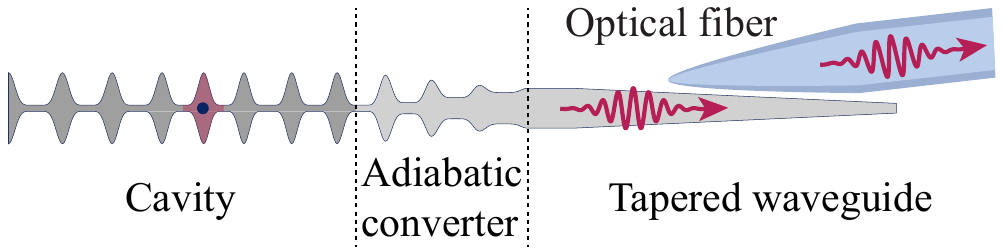}
\caption{Conceptual representation of the full Sawfish spin-photon interface. Due to the Purcell effect, a single quantum emitter (blue dot) positioned at the center of cavity defect (red area) emits photons resonant with the cavity mode at a higher rate. Such photons exit the cavity through an adiabatic converter and are efficiently coupled into a single-mode optical fiber via an optimized tapered outcoupler.}
\label{fig:f1_concept}
\end{center}
\end{figure}

Two main directions are currently pursued for the realization of free-standing nanostructures in bulk diamond, which are both based on plasma etching techniques: the Faraday-cage assisted angled etching~\cite{Burek2012, Bayn2014, Latawiec2016a} and the quasi-isotropic plasma etching~\cite{Khanaliloo2015a, Khanaliloo2015, Mouradian2017, Wan2018, Rugar2021, Kuruma2021}. In this work, we apply the latter procedure for the fabrication of Sawfish cavities with embedded NV$^{-}$ color centers. We find that our method enables the fabrication of devices with a wide range of parameters across the whole chip. Furthermore, we demonstrate that the Sawfish design is robust against structural imperfections and allows for the precise estimation of the cavity resonance wavelength from a simple inspection of the structure in a scanning electron microscope (SEM).

\section{The Sawfish cavity design}

\begin{figure}[t!]
\begin{center}
\includegraphics[width=8.5cm]{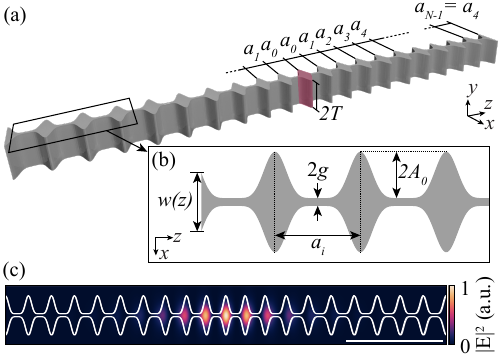}
\caption{Sawfish photonic crystal cavity design. (a) Schematic illustration of the Sawfish cavities presented in this work. The cavity defect (marked in red) is situated at the center of a suspended nanobeam of thickness $2T$ and is formed by 4 unit cells, each of which has a lattice constant increasing from $a_0$ to $a_3$. $N-4$ additional unit cells, with lattice constant $a_4$, define the mirrors on both sides of the cavity. (b) Visualization of the characteristic parameters of the Sawfish unit cell. The width $w(z)$ of the device varies along the propagation direction $z$ according to Eq.\ref{eq:Sawfish} in the main text. The corrugation has amplitude $2A_0$ and occurs at every lattice constant $a_i$. The central vein of the structure has width $2g$. (c) Electric field intensity $|E|^2$ of the TM-like mode at $\lambda=\SI{619}{\nano\meter}$ for a Sawfish cavity designed with the following set of parameters: $[a_0,a_1,a_2,a_3,a_4]=[200.0,207.0,217.9,228.4,238.5]~\SI{}{\nano\meter}$, $T=\SI{133}{\nano\meter}$, $N=10$, $A_0=\SI{65}{\nano\meter}$ and $g=\SI{11}{\nano\meter}$. The scale bar is $\SI{1}{\micro\meter}$ long.}
\label{fig:f2_design}
\end{center}
\end{figure}

Fig.~\ref{fig:f1_concept} shows a conceptual representation of the full fiber-coupled Sawfish cavity design. Three distinct parts can be identified: on one side, a color center in diamond is spatially and spectrally coupled to the Sawfish PhCC and constitutes the efficient light-matter interface that generates highly coherent photons. On the other side, an adiabatically tapered outcoupler channels the emitted photons into a single-mode optical fiber and links the device to the rest of the quantum network. These two sections are connected by an adiabatic converter, where the Bloch mode of the cavity is converted into the one supported by the rectangular waveguide. In the present work, we focus on the fabrication and characterization of the cavity section, which is schematically illustrated in Fig.~\ref{fig:f2_design}.

The device consists of a diamond nanobeam with thickness $2T$ and width $w(z)$ that periodically varies along the propagation direction $z$ according to the following function~\cite{Bopp2022} 
\begin{equation}
	w(z)=2\cdot\left[2A_0\cos^{6}\left(\frac{\pi}{a_i}\cdot z\right)+g\right].
	\label{eq:Sawfish}
\end{equation}
Here, $2A_0$ is the maximal amplitude of the oscillation, $a_i$ is the lattice constant of the photonic crystal's i-th unit cell and $g$ is the half-width of the central vein (Fig.~\ref{fig:f2_design}b). The central defect that supports the cavity mode (marked red in Fig.~\ref{fig:f2_design}a) is defined by progressively increasing the value of the lattice constant on both sides of it, from $a_0$ to $a_3$. The remaining $N-4$ unit cells of the photonic crystal have lattice constant $a_4$ and constitute the outer mirrors of the cavity. Fig.~\ref{fig:f2_design}c shows the simulated electric field intensity of the TM-like cavity mode for a Sawfish cavity that has been optimized for the ZPL emission of SnV$^-$ in diamond at $\lambda=\SI{619}{\nano\meter}$.
\begin{figure}[t!]
\begin{center}
\includegraphics[width=8.5cm]{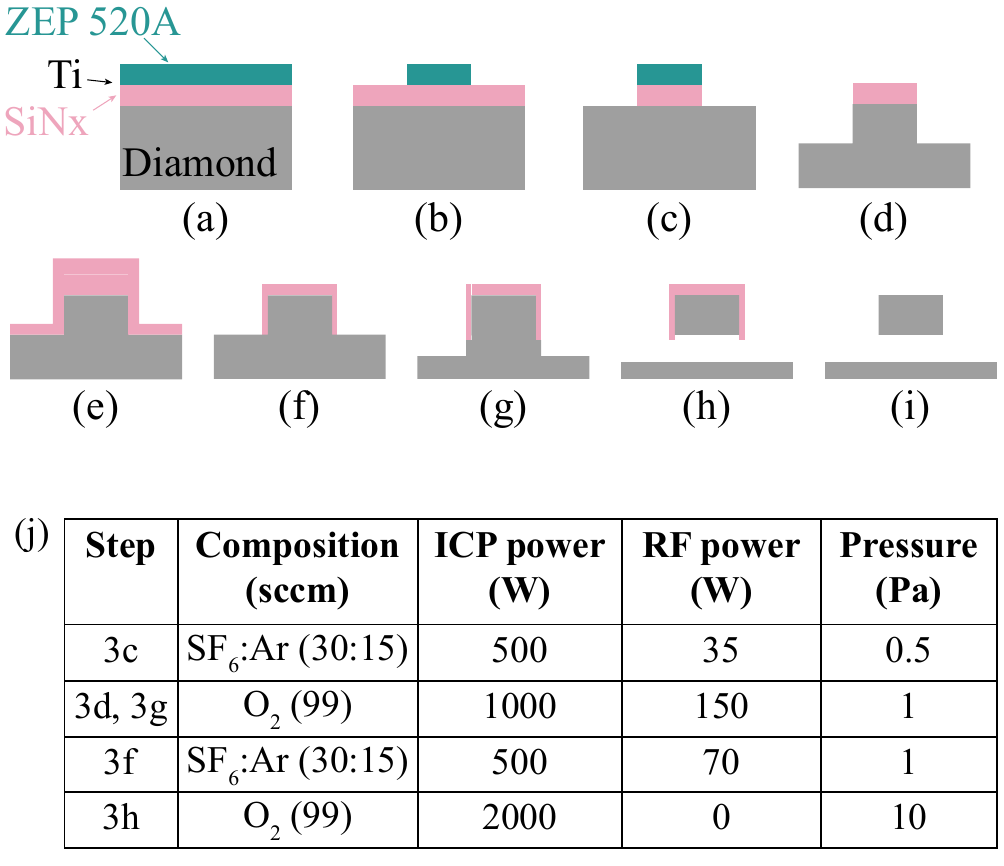}
\caption{(a)-(i) Schematic of the fabrication steps. (a) The etch mask consists of a 200-nm-thick layer of SiN$_x$ and \SI{10}{\nano\meter} of Ti. The stack of layers is coated with about \SI{300}{\nano\meter} of e-beam resist (ZEP520A). (b) The pattern of the cavities is transferred into the diamond via e-beam lithography, followed by (c) SF$_6$-plasma etch step to open the hard mask and (d) anisotropic O$_2$-plasma to pattern the diamond. (e) A second layer of SiN$_x$ is conformally deposited on the nanostructures, in order to protect the sidewalls. (f) A second anisotropic etch with SF$_6$-plasma removes the SiN$_x$ that covers the substrate, while the thickness of the vertical surfaces is reduced only marginally. (g) An additional anisotropic O$_2$-etch is performed to expose the sidewalls at the bottom of the nanostructures. This increases the amount of diamond that interacts with the O$_2$ plasma during the quasi-isotropic etching (h), thus facilitating the release of the cavities from the substrate. (i) Finally, a dip in buffered HF dissolves the remaining SiN$_x$ and uncovers the protected surfaces of the diamond structures. (j) Nominal parameters for the different plasma-etching recipes used during the steps shown in the figure.}
\label{fig:f3_fab}
\end{center}
\end{figure}

\section{Fabrication methods}
The substrate chosen for this work is a 3x3x\SI{0.25}{\milli\meter\cubed} optical grade, single-crystal diamond grown by chemical vapor deposition (supplied by Element Six Technologies Ltd. (UK) and with \{100\} faces). The high density of native NV$^-$ color centers in optical grade diamonds provides a natural abundance of emitters already embedded inside the fabricated nanostructures, thus facilitating their characterization. We therefore adjust the original design of the cavities (cf. Fig.~\ref{fig:f2_design}c) to target a resonance mode at $\lambda=\SI{637}{\nano\meter}$, corresponding to the ZPL emission of NV$^-$. From Finite-Element Method (FEM) simulations, we obtain the following parameters: $[a_0,a_1,a_2,a_3,a_4]=[200.0,207.5,218.8,229.3,238.4]~\SI{}{\nano\meter}$, $T=\SI{147.7}{\nano\meter}$, $N=22$, $A_0=\SI{65}{\nano\meter}$ and $g=\SI{11}{\nano\meter}$. We fabricate 108 Sawfish cavities, distributed in two identical matrices that present values of $A_0$ and $g$ ranging from \SI{50}{\nano\meter} to \SI{75}{\nano\meter} and from \SI{10}{\nano\meter} to \SI{50}{\nano\meter}, respectively, in order to compensate for possible fabrication-related imperfections and to demonstrate a wide tuning range for a constant thickness.

Fig.~\ref{fig:f3_fab} outlines the steps for the fabrication process. First, the diamond substrate is coated with a 200-nm-thick layer of SiN$_x$, which is deposited by means of Inductively Coupled-Plasma Enhanced Chemical Vapor Deposition (IC-PECVD). After the deposition of a Ti anti-charging layer, an electron-sensitive resist (ZEP520A) is spin-coated onto the surface of the sample (Fig.~\ref{fig:f3_fab}a). The Sawfish pattern is exposed via electron-beam lithography (Raith EBPG5200) and developed in ZED-N50 for \SI{90}{\second} at room temperature (Fig.~\ref{fig:f3_fab}b). The pattern is then transferred into the SiN$_x$ layer (Fig.~\ref{fig:f3_fab}c) by an ICP reactive ion etching process in a F-based plasma (see Fig.~\ref{fig:f3_fab}j for process parameters of the plasma-etching steps) and subsequently anisotropically etched into the diamond (Fig.~\ref{fig:f3_fab}d) in a O$_2$ plasma. To protect the exposed diamond sidewalls, \SI{200}{\nano\meter} of SiN$_x$ are conformally deposited by a second IC-PECVD process (Fig.~\ref{fig:f3_fab}e). A second etching in F-based plasma is then performed to expose again the diamond substrate, while only marginally reducing the thickness of the vertical surfaces (Fig.~\ref{fig:f3_fab}f). An additional anisotropic O$_2$-plasma etching step is carried out until a total depth of approximately \SI{1.2}{\micro\meter} is reached (Fig.~\ref{fig:f3_fab}g). This etch step exposes the diamond on the sidewalls of the nanostructures, which facilitates their release from the substrate. This is achieved in a quasi-isotropic etching step (Fig.~\ref{fig:f3_fab}h), where an O$_2$ ICP plasma with no RF power is employed until all the Sawfish cavities are fully suspended. A final dip in a buffered-HF solution completely dissolves the SiN$_x$ hard-mask layer, thus revealing all the diamond surfaces (Fig.~\ref{fig:f3_fab}i). 

\section{Characterization and discussion}
\begin{figure}[tb!]
\begin{center}
\includegraphics[width=8.5cm]{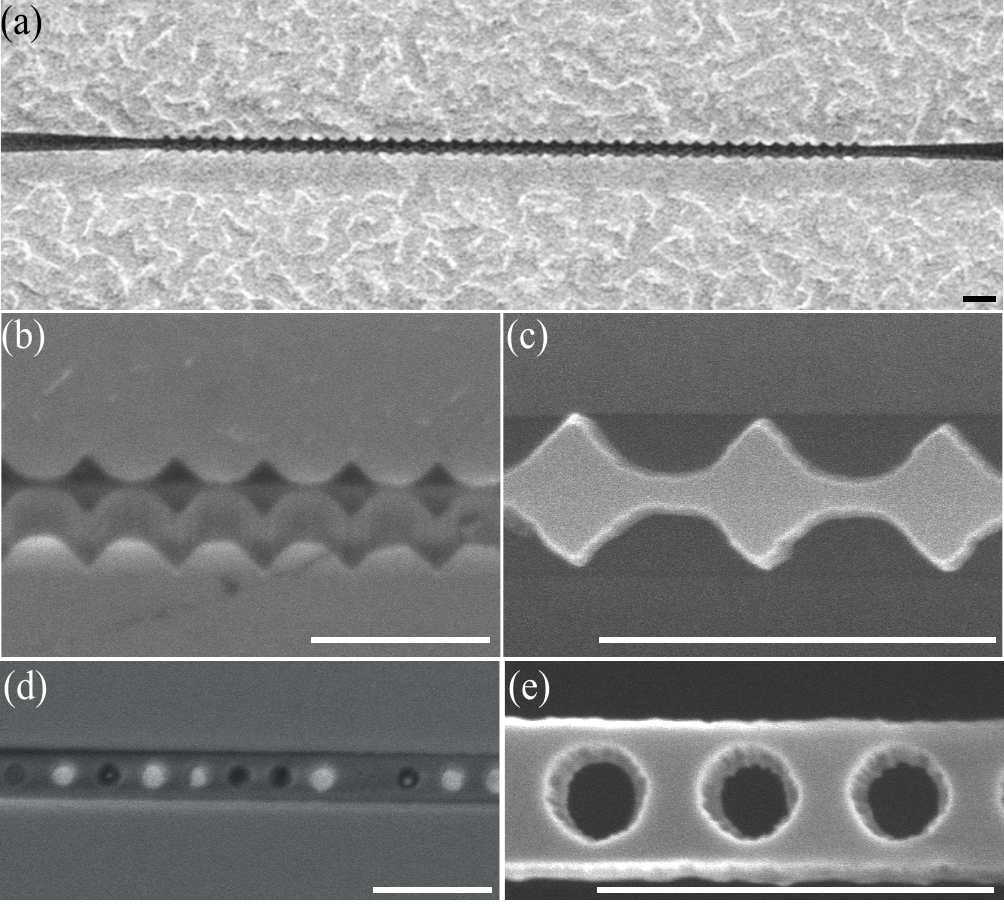}
\caption{SEM images of the fabricated devices. (a) Electron micrograph showing a \SI{30}{\degree}-tilted view of a fully suspended Sawfish cavity. (b) \SI{30}{\degree}-tilted and (c) top views of the width corrugation of the fabricated device, showing the roughness of the sidewalls. (d) SEM image of a hole-based PhCC fabricated on the same substrate as the presented Sawfish cavities for reference. The nominal hole radius is $r=\SI{60}{\nano\meter}$. Not all the holes are fully etched through. (e) Close-up SEM image of holes with nominal radius $r=\SI{64}{\nano\meter}$. The edges of the features present a pronounced roughness and the openings show sloped sidewalls. All scale bars correspond to \SI{500}{\nano\meter}.}
\label{fig:f4_SEM}
\end{center}
\end{figure}

The method described in the previous section allows for the successful fabrication of fully suspended Sawfish cavities with a total length of \SI{20.5}{\micro\meter}. Fig.~\ref{fig:f4_SEM} shows representative SEM images of the devices, which have a final thickness of $2T=\SI{275}{\nano\meter}$, as estimated from the \SI{30}{\degree}-tilted view shown in Fig.~\ref{fig:f4_SEM}b. The SEM inspection of the fabricated structures also reveals rough sidewalls and signs of erosion at the top edges of the devices, as visible in Figs.~\ref{fig:f4_SEM}b and c. We interpret these imperfections as an overetching effect of the hard-mask layer during the first F-based plasma step (Fig.~\ref{fig:f3_fab}c): due to low selectivity of the employed recipe, the resist mask gets increasingly damaged by the plasma and thus becomes a less effective protection for the SiN$_x$ layer underneath. The edges of the resist mask are particularly affected and they slowly overetch during this step, which causes the uncontrollable erosion of the hard mask at these locations. The damage is then transferred into the diamond layer during the subsequent etching step (Fig.~\ref{fig:f3_fab}d), resulting in the rough edges and sidewalls visible on the fabricated structures (Fig.~\ref{fig:f4_SEM}). In fact, narrow features are especially vulnerable to overetching, because such edge effects occur on both sides of the beam. Nonetheless, Sawfish cavities with $g$-values as small as \SI{20}{\nano\meter} were reliably fabricated. We believe that a thicker layer of ZEP resist may be more resilient against the plasma damage and would therefore improve the final structural quality of the fabricated devices.~\cite{Kuruma2023}

Furthermore, we note that hole-based PhCCs, which were fabricated on the same substrate as a reference, show holes that are consistently open only for nominal values of radius $r\geq\SI{64}{\nano\meter}$ (Fig.~\ref{fig:f4_SEM}d). The openings present high level of roughness and non-vertical sidewalls (Fig.~\ref{fig:f4_SEM}e). This comparison indicates that the Sawfish design requires less stringent fabrication parameters compared to the hole-based designs presented in literature.

\begin{figure}[t!]
\begin{center}
\includegraphics[width=8.5cm]{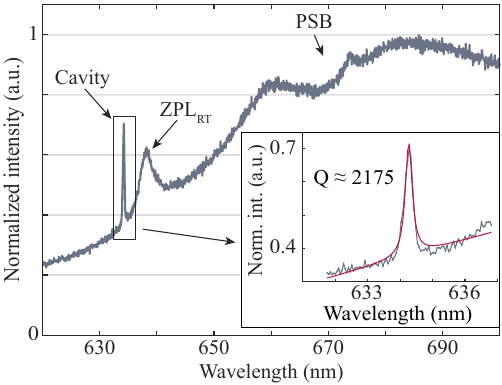}
\caption{Fluorescence spectrum for a cavity with $g=\SI{20}{\nano\meter}$ and $A_0=\SI{60}{\nano\meter}$, collected at room temperature. The typical spectral features of the NV$^-$ in diamond are recognizable, with a pronounced peak at about \SI{637}{\nano\meter}, corresponding to the ZPL emission, followed by the broad emission of the phonon sideband (PSB). The sharp peak on the left-hand side of the spectrum corresponds to the resonance mode of the cavity, which Purcell-enhances the NV$^-$ emission intensity in this narrow spectral band. Inset: Lorentzian fit of the cavity resonance, from which a Q-factor of approximately 2175 is estimated.}
\label{fig:f5_spectrum}
\end{center}
\end{figure}

We characterize the fabricated cavities by photoluminescence (PL) spectroscopy in two custom-made confocal microscopy setups operating at different temperatures: at room temperature (RT) and at T~=~\SI{4}{\kelvin}. In the former, the nanostructures are excited using approximately \SI{2}{\milli\watt} of a \SI{532}{\nano\meter} laser, which results in an above-band excitation of the native NV$^-$s embedded in the devices. The cavity emission is collected with a 0.95 NA objective and resolved in a spectrometer with a resolution of \SI{0.07}{\nano\meter}. The spectra at T~=~\SI{4}{\kelvin} are instead collected by using a continuous wave \SI{520}{\nano\meter} excitation laser set to \SI{200}{\micro\watt}. Here, the PL emission is recorded with a 0.82 NA cryo-objective and transferred via a single-mode fibre to a spectrometer with a resolution of \SI{0.06}{\nano\meter}. 

Fig.~\ref{fig:f5_spectrum} shows an example of the RT PL signal collected from the fabricated devices, where the typical spectral features of the NV$^-$ can be easily recognized: the broad ZPL peak at about \SI{637}{\nano\meter} next to the much wider emission caused by the phonon-sideband (PSB). The narrow emission that is blue-detuned from the ZPL peak corresponds to the resonance mode of the Sawfish cavity. No other peaks can be assigned to the cavity emission, which confirms the single-mode nature of the Sawfish design.~\cite{Bopp2022} 

\begin{figure}[tb!]
\begin{center}
\includegraphics[width=8.5cm]{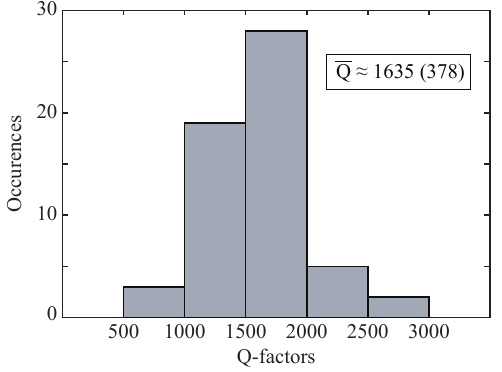}
\caption{Estimated values of Q-factors for the analyzed cavities at RT. The average $\overline{\text{Q}}$ of the distribution and the standard deviation are approximately 1635 and 378, respectively.}
\label{fig:f6_Qs}
\end{center}
\end{figure}

\begin{figure}[tb!]
\begin{center}
\includegraphics[width=8.5cm]{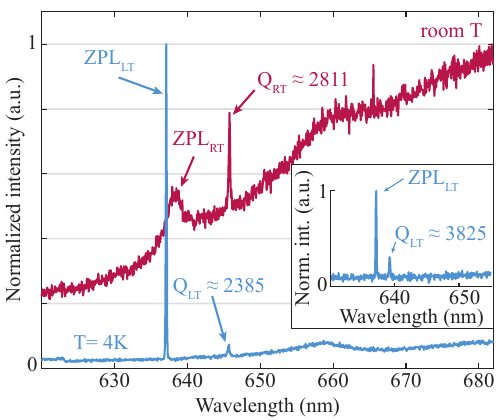}
\caption{Comparison between spectra collected at RT (red) and at \SI{4}{\kelvin} (cyan) for a cavity with $g=\SI{25}{\nano\meter}$ and $A_0=\SI{55}{\nano\meter}$. The low temperature quenches the phononic vibrations of the diamond lattice, which reduces the intensity of the PSB emission. A slight decrease of the estimated values of Q is detected. Inset: Spectrum collected at T~=~\SI{4}{\kelvin} for the outlier cavity, with $g=\SI{25}{\nano\meter}$ and $A_0=\SI{50}{\nano\meter}$, showing an increased Q-factor after cooling-down.}
\label{fig:f7_LT}
\end{center}
\end{figure}

A Lorentzian fit of the detected peaks allows for the estimation of the corresponding central wavelength and width and thus for the calculation of the Q-factors of the fabricated cavities. Fig.~\ref{fig:f6_Qs} shows the histogram of the estimated Qs, from which we calculate an average value and standard deviation of approximately 1635 and 378, respectively. The highest value was found to be $\text{Q}_{\text{RT}}\approx2811$. After cooling down the devices to \SI{4}{\kelvin}, we notice a general slight decrease of the  Q-factors (Fig.~\ref{fig:f7_LT}). We speculate that this change is probably due to condensation and freezing of water and other substances on the surface of the devices during the cooling-down process, resulting in an additional loss channel for the cavities. No significant changes can be detected for the spectral positions of the resonance modes. We note that one investigated cavity shows the opposite behavior, with a $\text{Q}_{\text{LT}}\approx3825$ at T~=~\SI{4}{\kelvin} from a $\text{Q}_{\text{RT}}\approx2633$ measured at room T (Fig.~\ref{fig:f7_LT}, inset). A thorough investigation of the cause behind this effect requires a more detailed study, which is however beyond the scope of the present work.
 
\begin{figure}[tb!]
\begin{center}
\includegraphics[width=8.5cm]{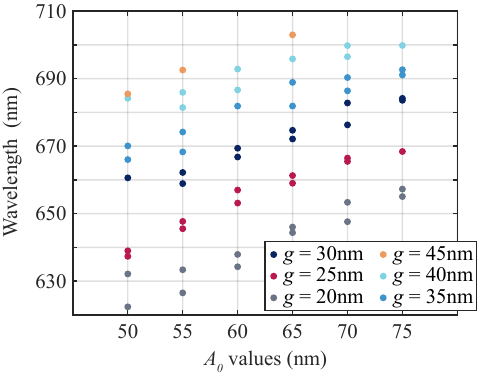}
\caption{Dependence of the experimentally estimated cavity resonances vs design parameters. As expected, the spectral positions of the cavity mode shift towards longer wavelengths for increasing values of $A_0$ and $g$. The uncertainties of the reported data points, obtained from the Lorentzian fit and calculated as half of the respective confidence intervals, are smaller than the displayed markers and are therefore not shown.}
\label{fig:f8_Freqs}
\end{center}
\end{figure}

In order to investigate the dependence between design parameters and spectral positions of the cavity resonances, in Fig.~\ref{fig:f8_Freqs} we report the wavelengths of the detected peaks as a function of the nominal values of parameters $A_0$ and $g$. As expected, the detected cavity modes shift towards longer wavelengths when either the central vein $g$ or the maximal amplitude of the corrugation $2A_0$ become wider. In fact, any increase in the values of the structural parameters defining the cavity (cf. Fig.~\ref{fig:f2_design}) results in a shift of the band structures towards lower frequencies.~\cite{Bopp2022} The consistency of such a trend for cavities fabricated with different $A_0$ and $g$ demonstrates that the presented fabrication method gives homogenous results over a wide range of parameters and for different devices across the chip. We note that no cavity modes could be detected in the spectra collected from devices with $g=\SI{50}{\nano\meter}$. This could be explained by the relation between the width of the central vein and the size of the photonic bandgap of the Sawfish cavity, which becomes narrower for increasing values of $g$.~\cite{Bopp2022} We conclude that the nanostructures showing no resonances are characterized by values of $g$ that are too large to generate efficient optical confinement. This results in a total of 72 devices that show both full structural integrity and the ability to host cavity resonances. With resonant modes detected in 57 nanostructures (cf. Fig.~\ref{fig:f8_Freqs}), we obtain a yield of about 79\% for the optically active resonators in the fabricated chip.

\begin{figure}[t!]
\begin{center}
\includegraphics[width=8.5cm]{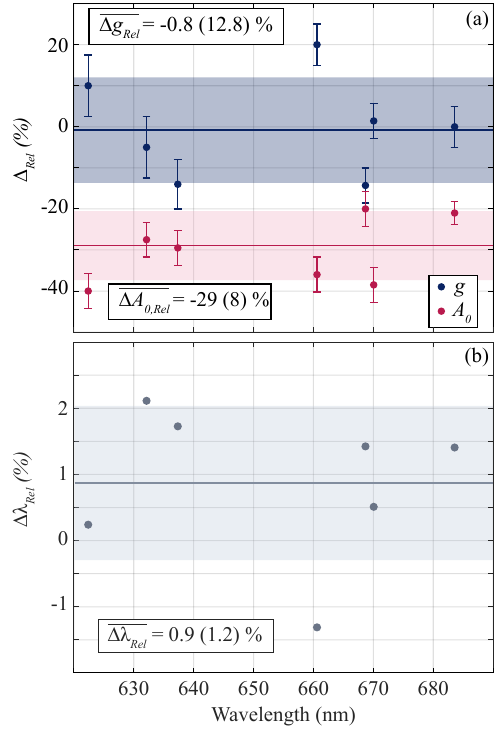}
\caption{(a) Distributions of the relative changes $\Delta g_{Rel}$ and $\Delta A_{0,Rel}$ (blue and red, respectively) from the designed values of $g$ and $A_0$ as measured from SEM images. The solid lines and the shaded areas correspond to the average values and the standard deviation, respectively. (b) Relative offset $\Delta\lambda_{Rel}$ between the measured cavity resonance emission and the value obtained by simulating Sawfish cavities designed by using values of $A_0$ and $g$ as measured from SEM images of corresponding nanostructures. On average, an offset of $0.9\%$ is obtained (solid line), with a calculated standard deviation of $1.2\%$ (visualized as light grey shaded area).}
\label{fig:f9_SimComp}
\end{center}
\end{figure}

Finally, we investigate the effects of the fabrication process on the Sawfish design. Fig.~\ref{fig:f9_SimComp}a plots the relative changes from the nominal values of $g$ and $A_0$ as measured from SEM images. The tips of the corrugation appear to be the features that are more affected, as we detect an average reduction of about $29\%$ from the designed values of $A_0$. The half-width of the central vein presents instead a smaller deviation, with an average of only $\overline{\Delta g_{Rel}}=-0.8\%$. This is to be expected because the tips of the corrugations are fine features that protrude into the wide trenches around the devices and are thus particularly exposed to the plasma; here, the soft resist mask is easily consumed and thus reduce its effectiveness. In contrast, the distribution of the values of $g$ has a larger standard deviation than the one for $A_0$ ($12.8\%$ vs $8\%$, respectively). This confirms that the aforementioned uncontrolled erosion of the edges has a larger impact on the narrow veins of the fabricated devices and makes their dimensions less homogenous across different devices. 

Fig.~\ref{fig:f9_SimComp}b shows the difference between the spectral positions of the detected resonance modes and those resulting from simulations of Sawfish cavities that were modeled by using the corresponding measured values of $g$, $A_0$ and thickness $2T$ (cf. Fig.~\ref{fig:f2_design}). Remarkably, the former differ, on average, by only 0.9\% from the simulated ones. This shows that, despite the unavoidable fabrication-related deviations from the expected design, a simple SEM inspection of the devices is enough to precisely estimate the resonance frequency of the cavities. Such a procedure enables the post-selection of only those nanostructures that are spectrally resonant to the embedded emitters, which is advantageous for the efficient assembly of large quantum photonics networks.

\section{Conclusions and outlook}

In summary, we have presented a method for the fabrication of our recently-proposed Sawfish photonic crystal cavity in bulk diamond. We successfully employed the described process and obtained fully suspended devices with a length of \SI{20.5}{\micro\meter} and features as narrow as \SI{40}{\nano\meter} across the full chip. The cavities consistently showed single resonance frequencies for a wide range of parameters, confirming their single-mode nature described by our simulations~\cite{Bopp2022} and demonstrating the robustness of the Sawfish design against the inevitable fabrication imperfections. This is further supported by a deviation of only 0.9 (1.2)\% between the measured resonance emissions and the values calculated by simulating Sawfish PhCCs with parameters extracted from the corresponding SEM images.

Future development will focus on improving the fabrication procedures in order to reduce the random erosion of the edges and thus increasing the cavity Q-factors. A better control over the overall process will also allow for the reproducible and reliable fabrication of Sawfish PhCCs that spectrally couple to the ZPL of individual color centers. Combined with the accurate positioning of a single emitter inside the cavity defect, which is possible with state-of-art localization approaches,~\cite{Pregnolato2020, Liu2021} the Sawfish PhCC will be able to achieve the high level of Purcell enhancement necessary for the efficient emission of coherent photons from color centers in diamond. With the addition of the optimized converter and the tapered outcoupler sections (sketched in Fig.~\ref{fig:f1_concept}, but not discussed in this work), the Sawfish design will constitute a valuable building block for the development of next-generation quantum-photonic networks and the generation of photonic resource states for quantum information processing.~\cite{Borregaard2020}

\begin{acknowledgments}
The authors gratefully acknowledge Ina Ostermay and Ralph-Stephan Unger for fruitful discussions. Funding for this project was provided by the European Research Council (ERC, Starting Grant project QUREP, No.~851810), the German Federal Ministry of Education and Research (BMBF, project DiNOQuant, No.~13N14921; project QPIS, No.~16KISQ032K; project QPIC-1, No.~13N15858; project QR.X, No.~KIS6QK4001) and the Einstein Foundation Berlin (Einstein Research Unit on Quantum Devices)
\end{acknowledgments}

\section*{Author Declarations}
\subsection*{Conflict of Interest}
J.M.B and T.S. have a patent pending.

\subsection*{Author Contributions}
\textbf{Tommaso Pregnolato:} Conceptualization (supporting); Formal Analysis (lead); Project Administration (supporting); Resources (supporting); Supervision (equal); Visualization (lead); Writing - Original Draft Preparation (lead). \textbf{Marco E. Stucki:} Investigation (equal); Resources (lead); Visualization (supporting); Writing - Original Draft Preparation (supporting). \textbf{Julian M. Bopp:} Conceptualization (equal); Formal Analysis (supporting); Investigation (equal); Software (lead); Visualization (supporting); Writing - Original Draft Preparation (supporting). \textbf{Maarten H. v. d. Hoeven:} Investigation (equal); Writing - Original Draft Preparation (supporting). \textbf{Alok Gokhale:} Investigation (supporting); Writing - Original Draft Preparation (supporting). \textbf{Olaf Kr\"{u}ger:} Resources (supporting); Supervision (supporting); Writing - Original Draft Preparation (supporting). \textbf{Tim Schr\"{o}der:} Conceptualization (equal); Funding Acquisition (lead); Project Administration (lead); Supervision (equal); Writing - Original Draft Preparation (supporting).

\section*{Data Availability}
The data that support the findings of this study are available from the corresponding author upon reasonable request.

\nocite{*}
\bibliography{SawFishFab} 

\end{document}